
\input harvmac.tex

\let\bmx=\bordermatrix
\let\bar=\overline
\def\nl{\hfill\break}
\let\<=\langle
\let\>=\rangle
\let\e=\epsilon
\let\ep=\epsilon
\let\tht=\theta
\let\t=\theta

\let\p=\prime

\noblackbox
\pretolerance=750
\lref\fs{P. Fendley and H. Saleur, Nucl. Phys. B388 (1992) 609}
\lref\YY{C.N. Yang and C.P. Yang, Phys. Rev. 150 (1966) 321, 327;
151 (1966) 258}
\lref\pkmag{P. Fendley and K. Intriligator, ``Central charges without
finite-size effects'' (hep-th/9307101) RU-93-26}
\lref\rZamc{A.B. Zamolodchikov, JETP Lett. 43 (1986) 730}
\lref\witti{E. Witten, Nucl. Phys. B202 (1982) 253}
\lref\VV{C. Vafa and E. Verlinde, work in progress}
\lref\sw{A. Schellekens and N. Warner, Phys. Lett. 177B (1986) 317,
Phys. Lett. 181B (1986) 339; K. Pilch, A. Schellekens, and N. Warner, Nucl.
Phys. B287 (1987) 317}
\lref\witell{E. Witten, Comm. Math. Phys. 109 (1987) 525}
\lref\withole{E. Witten, Phys. Rev. D44 (1991) 314}
\lref\rZandZ{A.B. Zamolodchikov and Al.B. Zamolodchikov, Ann.
Phys. 120 (1980) 253}
\lref\rAZi{A.B. Zamolodchikov, Adv. Stud. Pure Math. 19 (1989) 1}
\lref\tbaref{Al.B. Zamolodchikov, Nucl. Phys. B342 (1990) 695}
\lref\rAlZii{Al.B. Zamolodchikov, Nucl. Phys. B358 (1991) 497}
\lref\lvw{W. Lerche, C. Vafa, N. Warner, Nucl.  Phys. B324 (1989)
427}
\lref\MukVaf{S. Mukhi and C. Vafa, ``Two Dimensional Black-hole as a
Topological Coset Model of c=1 String Theory''
(hep-th/9301083), HUTP-93/A002}
\lref\rAlZiv{Al.B. Zamolodchikov, ``Resonance factorized
scattering and roaming trajectories'', Ecole Normale preprint ENS-LPS-355}
\lref\susy{S.J. Gates {\it et al},
{\it Superspace: One Thousand and One Lessons in
Supersymmetry} (Benjamin-Cummings, Reading, 1983) }
\lref\ZamoGold{Al.B. Zamolodchikov, Nucl. Phys. B358 (1991) 524}
\lref\witlg{E. Witten, ``On the Landau-Ginzburg description of $N$=2
minimal models'' (hep-th/9304026) IASSNS-HEP-93/10}
\lref\dy{P. DiFrancesco and S. Yankielowicz, ``Ramond Sectors and N=2
Landau-Ginzburg Models'', (hepth/9305037) SPhT 93/049}
\lref\rZandZ{A.B. Zamolodchikov and Al.B. Zamolodchikov, Ann.
Phys. 120 (1980) 253}
\lref\rAZi{A.B. Zamolodchikov, Adv. Stud. Pure Math. 19 (1989) 1}
\lref\tbaref{Al.B. Zamolodchikov, Nucl. Phys. B342 (1990) 695}
\lref\RSOS{A.B. Zamolodchikov, Landau Institute preprint,
September 1989}
\lref\ku{K. Kobayashi and T. Uematsu, Phys. Lett. B275 (1992) 361}
\lref\BCN{H.W.J. Bl\"ote, J.L. Cardy and M.P. Nightingale, Phys. Rev.
Lett. 56 (1986) 742; I. Affleck, Phys. Rev. Lett. 56 (1986) 746}
\lref\jnw{G. Japaridze, A. Nersesyan and P. Wiegmann,
Nucl. Phys. B230 [FS10] (1984)
511; P. Wiegmann, Phys. Lett. B152 (1985) 209}
\lref\hasen{
P. Hasenfratz, M. Maggiore and F. Niedermayer, Phys. Lett. B245 (1990) 522;
P. Hasenfratz and F. Niedermayer, Phys. Lett. B245 (1990) 529}
\lref\sausage{V.A. Fateev, E. Onofri and Al. B. Zamolodchikov, ``The
sausage model (integrable deformations of O(3) sigma model)'', LPTHE
preprint 92-46}
\lref\smirff{F. Smirnov, {\it Form-factors in Completely Integrable Models of
Quantum Field Theory},  Singapore: World Scientific (1992)}
\lref\cfiv{S. Cecotti, P. Fendley, K. Intriligator, and C. Vafa, Nucl.
Phys. B386 (1992) 405}
\lref\FSZii{P. Fendley, H. Saleur and Al.B. Zamolodchikov,
``Massless Flows II: the exact $S$-matrix approach'' (hepth/9304051) to
appear in Int. J. Mod. Phys. A}
\lref\afl{N. Andrei, K. Furuya, and J. Lowenstein, Rev. Mod. Phys. 55
(1983) 331}
\lref\cfiv{S. Cecotti, P. Fendley, K. Intriligator, and C. Vafa, Nucl.
Phys. B386 (1992) 405}
\lref\Zamsm{A. Zamolodchikov and Al. Zamolodchikov, Nucl. Phys. B379
(1992) 602}
\lref\ft{L.D Faddeev, L.A. Takhtajan, Phys.
Lett. A85 (1981) 375}
\lref\pkondo{P. Fendley, ``Kinks in the Kondo problem''
(cond-mat\-@babbage.sissa.it/ 9304031) BU-93-10}
\lref\lc{A. Ludwig and J. Cardy, Nucl.  Phys. B285 (1987) 687}
\lref\cvetkut{M. Cvetic and D. Kutasov, Phys. Lett. B240 (1990) 61}
\lref\evanshollo{J. Evans and T. Hollowood, Phys. Lett. B293 (1992) 100}

\Title{\vbox{\baselineskip12pt\hbox{BUHEP-93-17}\hbox{RU-93-29}}}
{\vbox{\centerline{Exact $N$=2 Landau-Ginzburg Flows}}}
\vglue .5cm
\centerline{Paul Fendley$^1$\footnote{$^\dagger$}
{address after Sep.\ 1, 1993: Department of Physics,
University of Southern California, Los Angeles, CA 90089-0484}
 and Ken Intriligator$^2$}
\vglue .5cm
\vglue .3cm
\centerline{$^1 $Department of Physics, Boston University}
\centerline{590 Commonwealth Avenue, Boston, MA 02215, USA}
\centerline{\it fendley@ryan.bu.edu}
\vglue .3cm
\centerline{$^2$ Department of Physics}
\centerline{Rutgers University, Piscataway, NJ 08855, USA}
\centerline{\it keni@physics.rutgers.edu}
\vglue 1.0cm
We find exactly solvable $N$=2-supersymmetric flows whose infrared fixed
points are the $N$=2 minimal models.  The exact $S$-matrices and the Casimir
energy (a $c$-function) are determined along the entire renormalization group
trajectory. The $c$-functions run from $c$=3 (asymptotically) to the $N$=2
minimal-model values, leading us to interpret these theories as the
Landau-Ginzburg models with superpotential $X^{k+2}$.  The calculation of the
elliptic genus is consistent with this interpretation. We also find an
integrable model in this hierarchy with spontaneously-broken supersymmetry and
superpotential $X$, and a series of integrable models with $(0,2)$
supersymmetry.  The flows exhibit interesting behavior in the UV, including a
relation to the $N$=2 super sine-Gordon model.  We speculate about the
relation between the kinetic term and the cigar target-space metric.
\medskip
\Date{7/93}
\newsec{Introduction}

\nref\rLG{A.B. Zamolodchikov Sov. J. Nucl. Phys. 44 (1986)
529;\nl D. Kastor, E. Martinec and S. Shenker, Nucl. Phys. B316 (1989) 590}
\nref\rVW{E. Martinec, Phys. Lett. 217B (1989) 431;\nl
C. Vafa and N.P. Warner, Phys. Lett. 218B (1989) 51}
\nref\strvac{B. Greene, C. Vafa, and N.P. Warner, Nucl. Phys. B324
(1989) 427;\nl C. Vafa, Mod. Phys. Lett. A4 (1989) 1169;\nl K. Intriligator
and C. Vafa, Nucl. Phys. B332 (1990) 95}
\nref\fmvw{P. Fendley, S.D. Mathur, C. Vafa and N.P. Warner,
Phys. Lett. B243 (1990) 257}
\nref\lw{W. Lerche and N. Warner, Nucl. Phys. B358 (1991) 571}
\nref\pk{P. Fendley and K. Intriligator, Nucl. Phys. B372 (1992) 533}
\nref\pkii{P. Fendley and K. Intriligator, Nucl. Phys. B380 (1992) 265}
\nref\lnw{A. LeClair, D. Nemeschansky and N. Warner, Nucl. Phys. B390 (1993)
653.}
The Landau-Ginzburg description of $N$=2 theories \refs{\rLG, \rVW} starts
with a $N$=2 superspace action written in terms of a chiral superfield (or,
more generally, several chiral superfields) as
\eqn\LGaction{S=\int d^2z d^4\Theta\ K(X, \bar X)+\int d^2z d^2\Theta\
W(X)+h.c.,}
where, for example, $K=\bar XX$ gives the standard flat-space kinetic terms.
This lagrangian has IR difficulties and appears to be intractable.  We do
know, however, that the superpotential $W$ is not renormalized (other than by
wavefunction renormalization) along the renormalization group flow of this
theory into the infrared.  This is the $N$=2 nonrenormalization theorem. By
dimension counting, changing the kinetic term $K$ amounts to an irrelevant
perturbation so an IR fixed point should be completely determined by the
superpotential $W$.  IR fixed points of the RG flow are $N$=2 superconformal
theories, so the LG approach allows us to describe many such theories simply
in terms of a superpotential.  This is why the LG description of $N$=2
theories is so useful in spite of the fact that directly analyzing the theory
along the flow is impossible at the present time.  A wide variety of $N$=2
theories can be described in this way, including the $N$=2 minimal models
\rVW\ and, via orbifolds, superstring vacua \strvac .
The power of the LG description also is applicable to perturbed $N$=2
superconformal theories \refs{\fmvw -\lnw }.

Given the success of the LG description it is natural to want to understand at
a deeper level the flow of \LGaction\ into its IR fixed point.  In this paper
we present integrable scattering theories which flow into the $N$=2 minimal
models.  We argue that these theories exactly describe the LG models along
their entire renormalization group trajectories.  These scattering theories
are described in terms of exact $S$-matrices for the scattering of the
massless excitations which survive in the IR limit.  Because this is an exact
and completely non-perturbative description of these quantum field theories,
it is a great challenge to connect these results with a direct analysis in
terms of the RG flow of the action \LGaction .  The hope is that these exact
results could lead to new insight into $N$=2 LG theories and quantum field
theory in general. For example, from an exact $S$-matrix one can compute exact
form factors \smirff, which can then be used to extact correlators.

Because of the difficultity in obtaining a direct connection between our exact
scattering theories and the RG flow of a LG action into the $N$=2 minimal
models, we will provide some (highly non-trivial) checks.  By analyzing the
thermodynamics of the scattering theories we will exactly determine the
Casimir energy (a $c$-function \refs{\rZamc,\lc}) along the entire RG
trajectory in terms of a solution of coupled (TBA) integral equations.  We
verify in sect.\ 2 that this $c$-function does flow from $c=3$
(asymptotically) in the UV to the minimal model value of the central charge in
the IR, as would be expected for the LG theory.

Because our $N$=2 scattering theories have individually-conserved left and
right fermion numbers $F_L$ and $F_R$, a second check on our scattering
matrices is to evaluate their elliptic genera \refs{\sw,\witell}
\eqn\elliptic{Tr\ e^{i\alpha _L F_L}(-1)^{F_R}q^{H_L}\bar q ^{H_R}.}
The fact that this quantity only receives contributions from states with
$H_R$=0 is nicely exhibited in our scattering theories. The insertion of
$(-1)^{F_R}$ causes the right-moving particles to decouple; only left
movers contribute to \elliptic\ all along the RG flow, all the way from the
far UV to the far IR.  We see that \elliptic , as computed in our
scattering theories, is constant along the RG flow. This behavior is required
for \elliptic\ because it is an index.  We derive the leading term (in the
thermodynamic limit $q\to 1$) from our scattering theories and verify that it
agrees with this limit of the exact expression, which was recently computed
using the $N$=2 LG field theory in \witlg\ and verified in terms of the known
$N$=2 minimal model characters in \dy .

In sect.\ 4 we discuss two related $S$-matrices. The first describes the
scattering of Goldstinos resulting from spontaneously-broken $N$=2
supersymmetry. We argue for a LG potential which describes this case and which
fits in nicely into our overall picture. This result also indicates that the
potential $K$ is not the simple one $\bar X X$.  The second model has $(0,2)$
supersymmetry and might be of interest for string theory model-building.

We then attempt to understand a little more about these LG theories,
especially their ultraviolet fixed points, whose properties are somewhat
confusing.  In sect.\ 5 and 6 we calculate power-series expansions of of the
$c$-function and the ground-state energy in an external background field
around the UV fixed point; we find that the results are the same as those of
the $N$=2 super-sine-Gordon model except for some alternating signs.  In
sect.\ 7 we speculate about comparing the UV limit of our exact results with
the LG theory \LGaction\ whose kinetic term $K$ corresponds to a cigar metric.

\newsec{Exact results for $N$=2 Landau-Ginzburg flows}

We study integrable RG flows whose IR limit is a non-trivial conformal field
theory. The basic idea is that the theories can be described by massless
excitations for which we can find an exact $S$-matrix.  In particular, we have
a set of left movers and a set of right movers and $S$-matrix elements
describing left-left, right-right, and left-right interactions.  Because of
the left-right interaction the massless theory is not conformal.  In the IR
limit of the RG flow, the left-right $S$-matrix becomes trivial, the left and
right movers decouple from each other, and we obtain the IR fixed point
conformal theory.  The left-left and right-right $S$-matrices are independent
of the RG scale and encode a description of the IR fixed point.  Such massless
scattering theories arise, for example, in the continuum limit of the XXX spin
chain \refs{\afl,\ft} and have recently been proposed for flows from the
$p^{\hbox{th}}$ to $(p-1)^{\hbox{th}}$ $N$=0 minimal models
\refs{\ZamoGold,\FSZii}, flows from the $SU(2)$ principal chiral model with a
WZW term into the $SU(2)_1$ CFT
\Zamsm, deformations of the O(3) sigma model with topological term
$\theta$=$\pi$
\refs{\Zamsm ,\sausage} and for the Kondo problem \pkondo.

The theories which we will be considering here are the $N$=2 minimal models
described by the superpotential \rVW
\eqn\Wis{W=gX^{k+2},}
where $g$ is a coupling constant.  We do not have a direct argument that such
flows should be integrable. However, it was shown in \witlg\ that, even in the
off-critical LG theory, there is a full $N$=2 superconformal algebra acting on
the pure left-moving states, i.e. those with $H_R$=0, and likewise for the
pure right moving states.  Since this is an infinite-dimensional symmetry
algebra, it is plausible that it leads to the conserved currents required
for integrability. We will assume integrability and find the simplest massless
scattering theories consistent with the symmetries and the various
requirements which exact $S$-matrices must satisfy.  The final results provide
highly nontrivial checks on our assumptions.

All excitations are all massless, with $H=|P|$.  The left movers form
representations of the left-moving $N$=2 supersymmetry algebra and are
annihilated by the right-moving generators.  In particular, the simplest
representation is a doublet $(u_L(\t ), d_L(\t ))$ under the action of the
generators $Q^{\pm}_L$:
$$Q^-_L|u_L(\tht)\rangle =\sqrt{2M}e^{-\tht/2}|d_L(\tht)\rangle\qquad\qquad
Q^+_L|d_L(\tht)\rangle=\sqrt{2M}e^{-\tht/2}|u_L(\tht)\rangle.$$
These excitations are eigenstates of $H_L=H-P=\{Q_L^+,Q_L^-\}$ with eigenvalue
$2Me^{-\t}$, where $M$ is a scale parameter, and have left-moving fermion
number $F_L$ given by $(f,f-1)$ for some $f$.  Likewise, the right-moving
excitations form doublets under $Q^{\pm}_R$, are eigenstates of $H_R=H+P$ with
eigenvalue $2Me^{\t}$, and are annihilated by the left-moving generators.
These are eigenstates of the right fermion number $F_R$.

The simplest scattering theory consists of a single left $(u_L,d_L)$ doublet
and a single right doublet.  The fermion numbers of the doublets are $(\half,
-\half)$ and the $d$ excitations are the anti-particles of the $u$
excitations.  Consider first the left-left $S$-matrix $S_{LL}$ for scattering
of the $u_L(\t)$ and $d_L(\t)$ among themselves.  Demanding that this
$S$-matrix commutes with the supersymmetry generators along with the
requirements of crossing and unitarity, and the stipulation that there be no
extra bound states completely determines the matrix $S_{LL}$.  In fact, the
$S$-matrix is formally the same as that obtained in \pk\ for massive $N$=2
theories with a spontaneously broken ${\bf Z}_2$ symmetry. Denoting
$u_L(\t_1)$ by  $u_1$, $S_{LL}$ is
\eqn\Smat{\bmx{&u_2u_1 &d_2 u_1& u_2
d_1&d_2d_1\cr
u_1u_2 &Z(\t)&0&0&0\cr
u_1d_2 &0&-iZ(\t)\tanh{\t\over 2}&Z(\t){1\over \cosh{\t\over 2}}&0\cr
d_1u_2 &0&Z(\t){1\over \cosh{\t\over 2}}&-iZ(\t)\tanh{\t\over 2}&0\cr
d_1d_2 &0&0&0&Z(\t)\cr},}
where $\t=\t_1-\t_2$ and
$$Z(\t)=\exp\left({i\over 4}
\int_{-\infty}^{\infty} {d\omega\over \omega}
{\sin \omega\t \over \cosh ^2{\pi \omega\over 2}}\right).$$
The $S$-matrix $S_{RR}$ for right-right scattering is also given
by \Smat.

Without a left-right interaction, these $S$-matrices would describe an IR
fixed point CFT.  In order to describe a massless but not conformal flow, we
now consider coupling the above left and right scattering theories with some
nontrivial left-right $S$-matrix.  Consider $u_L(\t_1)u_R(\t_2)$ scattering.
By fermion-number conservation the final state must be
$A(\t)u_R(\t_2)u_L(\t_1)$ for some function $A(\t)$.  Since the $S$-matrix
must commute with the generators $Q^{\pm}_L$ and $Q^{\pm}_R$, we see that
the $S$-matrix for left-right scattering must be diagonal with all elements
equal to the same function $A(\t)$.  The simplest nontrivial choice for this
function which satisfies the left-right $S$-matrix requirements \ZamoGold\ is
\eqn\slr{S_{LR}(\t)=\tanh \left({\t\over 2}-i{\pi\over 4}\right).}
This $S$-matrix does indeed becomes trivial in the UV and IR fixed point
limits.

The above is the simplest possible scattering theory for a massless but not
conformal $N$=2 supersymmetric flow.  We conjecture that it describes the LG
flow associated with $k$=1 case of \Wis , i.e. the simplest, massless $N$=2 LG
flow.

As a highly nontrivial check on the LG conjecture, we calculate the partition
function on a torus with euclidean time $\beta$ and length $L$ from this
scattering theory. This is done by using the thermodynamic Bethe ansatz
\tbaref. We take the fermion boundary conditions to be periodic in the $L$
direction and we insert the operator $e^{i\alpha _LF_L}e^{i\alpha _R F_R}$ to
give a field with fermion numbers $(q_L,q_R)$ the twisted boundary conditions
$-e^{i(\alpha _Lq_L+\alpha _R q_R)}$ in the $\beta$ direction.  In the
thermodynamic limit $L\to \infty$, we derive exact integral equations for the
corresponding free energy.  We give the final result without elaboration
because the techniques involved were discussed at length in \pk; it is
\eqn\TBAFis{\eqalign{c(\alpha_L,\alpha_R;M\beta)&\equiv
{6\beta\over \pi L}\log\Tr e^{i\alpha_LF_L}e^{i\alpha _R
F_R}e^{-\beta H}\cr
&={3\over \pi^2}\sum _a
\int d\t\ \nu _a(\t)\log(1+\lambda _ae^{-\e _a(\t)}),\cr}}
where $\e_a(\t)$ are obtained by solving the coupled integral equations
\eqn\TBAe{\e_a (\t) =\nu _a(\t) -\sum_b l_{ab}\int {d\t '\over 2\pi}
{1\over \cosh (\t
- \t ')}\log(1+\lambda _ae^{-\e _a (\t ')}).}
The index $a$ as well as the $\nu_a (\t)$, the $\lambda _a$, and
$l_{ab}$ in these equations are conveniently encoded in the diagram:
\bigskip
\noindent
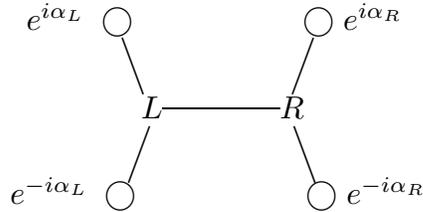
\centerline{
\hbox{\rlap{\raise28pt\hbox{$\hskip.3cm e^{i\alpha _L}\hskip.1cm\bigcirc\hskip
1.8cm\bigcirc\hskip.1cm e^{i\alpha _R}$}}
\rlap{\lower27pt\hbox{$e^{-i\alpha_L}\hskip.1cm\bigcirc\hskip
1.8cm\bigcirc\hskip.1cm e^{-i\alpha _R}$}}
\rlap{\raise14pt\hbox{$\hskip1.1cm\Big\backslash\hskip1.6cm\Big/$}}
\rlap{\lower14pt\hbox{$\hskip1.05cm\Big/\hskip1.55cm\Big\backslash$}}
$\hskip1.1cm${\hbox{$L$}}\raise.1cm\hbox{\vrule width1.3cm height
.4pt}{\hbox{$R$}} }}
\bigskip
\noindent
The index $a$ runs over each node in this diagram.  The $\nu _a(\t)$ are given
by zero if node $a$ is open, $\half M\beta e^{-\t}$ for the node with a $L$
and $\half M\beta e^{\t}$ for the node with a $R$ in it.  The $\lambda_a$ are
one except for the four outside nodes, for which the $\lambda _a$ are given by
the phases indicated in the diagram.  Finally, $l_{ab}=1$ if the nodes $a$ and
$b$ are connected by a line in the diagram and zero otherwise. This result
follows almost immediately from the results of \ZamoGold\ and \pk; the
coupling between the $L$ and $R$ nodes follows from the $S$-matrix \slr, while
the extra massless nodes arise from ``diagonalizing'' the $S$-matrix \Smat.

The above scattering theory and the resulting integral equations are to be
associated with the $k$=1 case of \Wis .  The $S$-matrices associated with the
higher $k$ cases are more complicated and we will not write them out
explicitly here.  They follow from the results of \FSZii, where analogous
scattering theories describing the flows between the $N$=0 minimal models are
discussed.  The spectrum in the $N$=0 case consists of massless solitons
interpolating between adjacent wells of a $\phi^{2(k+1)}$-type potential with
$k+1$ degenerate wells. In the $N$=2 case each particle merely becomes a $u,d$
doublet; the $LL$ and $RR$ scattering is then the $N$=0 $S$-matrix tensored
with \Smat, while $S_{LR}$ is the same in the $N$=0 and $N$=2 cases \foot{If
one replaces the $S$-matrix \Smat\ with the $SU(2)$ doublet $S$-matrix of
\Zamsm, this gives the $S$-matrix for the flows into $SU(2)_k$ WZW models.}.
The TBA integral equations for higher $k$ become an obvious generalization of
the above ones: they are given by the same expressions \TBAFis\ and \TBAe\ but
with the diagram generalized to
\bigskip
\noindent
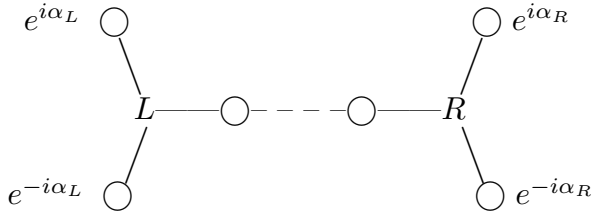
\centerline{
\hbox{\rlap{\raise28pt\hbox{$\hskip .3cm e^{i\alpha _L}\hskip.1cm\bigcirc\hskip
3.7cm\bigcirc\hskip.1cm e^{i\alpha _R}$}}
\rlap{\lower27pt\hbox{$e^{-i\alpha_L}\hskip.1cm\bigcirc\hskip
3.7cm\bigcirc\hskip.1cm e^{-i\alpha _R}$}}
\rlap{\raise14pt\hbox{$\hskip1.1cm\Big\backslash\hskip3.5cm\Big/$}}
\rlap{\lower14pt\hbox{$\hskip1.05cm\Big/\hskip3.45cm\Big\backslash$}}
$\hskip1.05cm${\hbox{$L$}}------$\bigcirc$-- --
-- --$\bigcirc$------$R$ }}
\bigskip
\noindent
where there are $k-1$ open nodes between the $L$ one and the $R$ one.

The reason for our normalization and notation in the expression \TBAFis\ is
that, by changing our interpretation about which cycle on the above torus is
length and which one is time, the free energy per unit length is proportional
to the ground-state (Casimir) energy of the quantum field theory on a circle
of radius $\beta$.  This is, in turn, proportional to the central charge at
the fixed points \BCN.  The function $c(\alpha _L, \alpha _R; M\beta )$ can be
interpreted as a $c$-function for the RG flow of the theory in the sector with
fermion boundary conditions twisted by $\alpha _L$ and $\alpha _R$.  $M\beta $
is the RG parameter which runs from zero in the far UV to infinity in the far
IR.  In these limits the function $c(\alpha _L, \alpha _R;
\beta M)$ coincides with the minimum value of $c-12(h+\bar h)$ in the fixed
point theory in the sector with the twisted fermion boundary conditions.

This allows a very convincing check that our scattering theories describe the
LG flow of \Wis\ because \TBAFis\ gives the correct central charge in both the
IR and the UV limits. Because our scattering theory is defined in terms of the
massless excitations associated with the IR limit of the RG flow, we first
verify that the IR limit of the TBA equations \TBAe\ do indeed correctly give
the central charge of the $N$=2 minimal models associated with the IR fixed
points of \Wis .  In the IR $\beta M\rightarrow \infty$ limit the integral
equations \TBAe\ effectively break up into a piece corresponding to the left
movers described by the diagram
\bigskip
\noindent
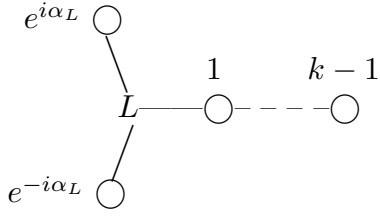
\centerline{
\hbox{\rlap{\raise28pt\hbox{$\hskip.3cm e^{i\alpha _L}\hskip.1cm\bigcirc$}}
\rlap{\lower27pt\hbox{$e^{-i\alpha_L}\hskip.1cm\bigcirc$}}
\rlap{\raise14pt\hbox{$\hskip.95cm\Big\backslash$}}
\rlap{\lower14pt\hbox{$\hskip.9cm\Big/$}}
\rlap{\raise12pt\hbox{$\hskip1.05cm \hskip .80cm 1\hskip .95cm k-1$}}
$\hskip.75cm${\hbox{$L$}}------$\bigcirc$-- --
-- --$\bigcirc$ }}
\bigskip

\noindent
and an analogous piece for the right movers.  We can evaluate the
corresponding leading contribution to \TBAFis\ exactly in terms of Rogers
dilogarithm functions using the standard trick \tbaref .  We omit the details,
as they closely follow calculations discussed in our previous papers.  The
final result is
\eqn\cir{c(\alpha _L, \alpha _R, \beta M\rightarrow \infty)={3k\over
k+2}(1-\half {\alpha _L^2\over\pi^2} - \half {\alpha _R^2\over \pi^2}).}
When $\alpha _L$=$\alpha _R$=0, i.e. in the NS sector, the minimum value
of $h+\bar h$ is zero, corresponding to the NS vacuum.  The result \cir\
then gives the correct value $c=3k/(k+2)$ for the central charge of the
$N$=2 minimal model associated with the IR fixed point of \Wis .  The
value of \cir\ for nonzero $\alpha _L$ and $\alpha _R$ is exactly as
expected for the sector with fermion boundary conditions twisted by
$\alpha _L $ and $\alpha _R$:  by spectral flow of the NS vacuum, the
minimum value of $h+\bar h$ in the sector with twisted fermion boundary
conditions is $c((\alpha _L/2\pi)^2+(\alpha _R/2\pi)^2)/6$, in agreement
with \cir .  This agreement is a check that the conserved charges $F_L$ and
$F_R$ associated with our scattering theory do, indeed, correspond to
left and right fermion number in the IR theory.

More support for the LG interpretation comes from finding the dimension of the
irrelevant operator which dominates the final stage of the RG flow into the
$N$=2 minimal model fixed point. This can be obtained from \TBAe\ either
numerically or by using the ``periodicity'' argument discussed in
\ZamoGold .  For $k\ne 1$ the dimension is $2+(4/k+2)$.
This shows that for our scattering theories the kinetic term in \LGaction\
flows as $K=K_{IR}+\lambda (X\bar X)^2+\cdots$, with $\lambda \to 0$ in the IR
limit. For $k$=1, the dimension is $3$, which corresponds to adding $(X\bar
X)^3$. Since the flow reaches the IR fixed point by a pure kinetic-term
perturbation, this is a good hint that it is purely LG kinetic-term all
the way as conjectured.

We now turn to the UV limit.  The $M\beta\rightarrow 0$ limit of \TBAe\ can,
again, be evaluated in terms of dilogarithm functions.  In particular, it is
seen for $\alpha _L$=$\alpha _R$=0, that $c(M\beta \rightarrow 0)$=3, for all
$k$.  This is precisely what we would expect for the UV limit of the LG
theory: the superpotential \Wis\ is driven to zero by wavefunction
renormalization and we are left with the central charge of a free superfield.
We should, however, expect subtleties.  By considering the Witten index or the
elliptic genus to be discussed in the next section, it is seen that the effect
of the superpotential can't completely go away, even in the far UV limit.  A
manifestation of this is that $c$=3 is only reached asymptotically.  The TBA
equations
\TBAe\ give
\eqn\cuv{c(M\beta \rightarrow 0)=3-{3\pi^2(k+2)\over 2(-\log M\beta +
const.)^2}+ \dots ,}
where the $\dots$ includes higher $1/\log(M\beta )$ terms as well as powers of
$M\beta$.  It is possible to derive this result from \TBAe\ along the lines of
the discussion in \sausage , but the calculation is subtle and we will not
present it here.  We verified this result by obtaining the numerical solution
of equations \TBAe\ for small $M\beta$; the results fit \cuv\ well.  The
$1/\log(M\beta)$ terms in \cuv\ reveal that $c$=3 is only approached
asymptotically.  For example, the derivative of the $c$-function blows up at
$M\beta$=0 so in this region the theory can not be thought of as a CFT with a
small perturbation.  We indeed see that the superpotential doesn't really go
away in the UV limit.  This is the statement that the UV fixed point is an
``infinite distance'' from the IR fixed point in whose neighborhood the
$S$-matrix is defined, an unavoidable consequence of the fact that the two
fixed points have different Witten indices \refs{\witti,\cvetkut}. In terms of
a LG theory \LGaction , one might also expect such log terms in the UV, coming
from unbounded fluctuations of the (near) constant modes of the bosonic
component of the superfield $X$.  In sect.\ 6 we will discuss these $1/\log$
terms in the context of a particular choice of the kinetic term $K$.

We also mention that the UV limit of \TBAe\ exhibits some interesting features
when $\alpha_L$ and $\alpha _R$ are turned on.  Because the UV region cannot
be described by a $c$=3 CFT plus a small perturbation, we shouldn't be
surprised to find that \TBAe\ differs in the UV from the result $3(1-\half
(\alpha _L/\pi)^2 -\half (\alpha _R/\pi)^2)$ of a $c$=3 CFT with boundary
conditions twisted by $\alpha_L$ and $\alpha _R$.  In particular, as will be
discussed in the following section, at $\alpha _R=\pi$ we are computing the
elliptic genus and we know there that \TBAe\ can't be of this form.
Especially interesting behavior occurs when $\alpha _L=\pm \alpha _R\equiv
\alpha $.  For example, for $k$=1 \TBAe\ yields
$c(\alpha;M\beta \rightarrow 0)=$
\eqn\cuvalpha{
\eqalign{
&3-5{\alpha ^2\over \pi ^2} +{f_1(\alpha)\over log^2M\beta}+\cdots
\qquad |\alpha |\leq {\pi \over 2}\cr
&(5-3{|\alpha |\over \pi})(1-{|\alpha |\over
\pi})f_2(\alpha) (M\beta )^{4(2|\alpha |-\pi)/3\pi}+\cdots
\qquad  {\pi \over 2}\leq |\alpha |\leq \pi \cr}}
It seems that at $\alpha=\pi/2$ the log correction turns into a power series
one with continuously varying exponent; the behavior is continuous in $\alpha$
but not analytic.  This reveals an interesting transition in $\alpha$,
possibly due to a level crossing associated with a state which would be
eliminated when $\alpha _L\neq \pm \alpha _R$ by the twisted boundary
conditions.

\newsec{The Elliptic Genus}

The elliptic genus \refs{\sw,\witell} of a supersymmetric theory having
separately-conserved left and right fermion numbers $F_L$ and $F_R$ is given by
\elliptic.  (If, as in heterotic theories, there is no $F_L$, the $e^{i\alpha
_LF_L}$ term should, of course, be omitted).  By standard arguments \witti ,
only those states with $H_R=0$ contribute to \elliptic\ and, thus, the
elliptic genus is independent of $\bar q$.  The elliptic genus is an ``index''
--- it is invariant under continuous deformations of the theory.

These features of the elliptic genus are nicely exhibited in our scattering
theory.  In particular, consider the TBA equations \TBAe\ with $\alpha
_R=\pi$.  Because these equations were obtained by taking the thermodynamic
limit where the length $L\rightarrow \infty$, they yield the $q=e^{-2\pi
\beta/L}\rightarrow 1$ limit of \elliptic .  The fact that the elliptic genus
is a holomorphic function of $q$, receiving contributions only from $H_R=0$
states, is reflected in the fact that the insertion of $(-1)^{F_R}$ causes the
right-moving excitations to decouple.  To see this from our integral equations
\TBAFis\ notice that a solution of
\TBAe\ is given by
\eqn\rdecouple{e^{-\e _R(\t)}=0, \qquad \e _{\pm F_R}(\t)=0,}
where $\e _R$ is the $\e _a$ for the node labeled by an $R$ and $\e_{\pm F_R}$
are the $\e_a$ for the two nodes labeled by $e^{\pm i\alpha _R}$ in the
diagram.  Therefore $\log (1+e^{-\e _R(\t)})$=0 for all $\t$ and thus the $R$
node as well as the two end nodes to the right of it are effectively cut off
of the diagram; we are left with the diagram appearing before \cir .  The
remaining system of integral equations is independent of the RG flow parameter
$M\beta$, as is to be expected since we are computing an index.  This is
easily seen by noting that a rescaling of $M\beta$ can now be absorbed into a
shift of $\t$ in \TBAFis\ and \TBAe .  The value of \TBAFis\ along the entire
RG flow is thus given by the IR expression \cir\ with $\alpha _R$= $\pi$; so
our value for the thermodynamic limit of \elliptic\ is
\eqn\thelliptic{Tre^{i\alpha _LF_L}(-1)^{F_R}e^{-\beta H}=\exp({\pi kL\over
4\beta (k+2)}(1-{\alpha_L^2\over \pi^2}))}
along the entire renormalization-group trajectory.  This agrees with the
$q\rightarrow 1$ limit of the exact $N$=2 minimal models result, recently
computed exactly using a free field theory associated with the LG description
in \witlg\ and verified in terms of the known minimal model characters in \dy.
Note that when $\alpha_L=\pm \pi$, \elliptic\ becomes equal to the Witten
index and the expression \thelliptic\ properly becomes independent of $\beta$
and $L$.

We also note that, as discussed in \cfiv , the thermodynamic limit of
Tr$F(-1)^Fe^{-\beta H}$ is a ``pseudo-topological'' index which is invariant
under pure ``D-term'' variations of the theory.  This index is especially
useful for analyzing massive theories, where the elliptic genus can not be
defined.  In the present context it is simply follows from the elliptic genus:
taking the derivative of \thelliptic\ with respect to $\alpha _L$ and setting
$\alpha _L$=$\pm \pi$ gives $TrF_L(-1)^{F_R\pm F_L}e^{-\beta H}$=$\mp {L\over
2\beta}(k/(k+2))$.  Adding or subtracting the same expression with $L$ and $R$
interchanged gives $TrF(-1)^Fe^{-\beta H}$ for the two choices of $F= F_L\pm
F_R$.  The result agrees with the minimal model result \cfiv .

\newsec{Related models: spontaneously-broken and (0,2) supersymmetry}

We briefly discuss two types of related models. The first model, that of
Goldstinos resulting from spontaneously-broken $N$=2 supersymmetry, in fact
fits in nicely with the above LG flows and gives a great deal of independent
support for the picture we have described.

\subsec{$N$=2 Goldstinos and the superpotential $W=X$}

If supersymmetry is spontaneously broken, one expects a massless fermionic
excitation for each broken generator. A simple and elegant example of this is
in the flow from the tricritical Ising model to the Ising model. The $N$=1
supersymmetry of the tricritical Ising model is spontaneously broken; in the
IR limit the resulting Goldstinos become the massless free Majorana fermion of
the Ising model \rLG. In the Landau-Ginzburg picture, this is described by
a supersymmetric $\phi^4$ Lagrangian; the effective Goldstino action is given
by integrating out the boson. These interactions are irrelevant so at the end
of the flow we are left with the free massless fermion, but in the midst of
the flow the fermions interact.  This flow is integrable, and as shown in
\ZamoGold, one can find the $S$-matrix for the left and right fermion:
$S_{LR}$ is given by \slr, while $S_{LL}$ and $S_{RR}$ must be $1$ because the
infrared limit is a free theory.

A virtually identical situation happens with $N$=2 supersymmetry. The only
difference is that because there are two supersymmetries, we will have two
left-moving particles with fermion number $F_L=\pm 1$ and likewise two right
movers with $F_R=\pm 1$. These can be thought of as a massless Dirac fermion.
In the IR limit, the fermion is free, so $S_{LL}=S_{RR}=1$. For the same
reasons as in the $N$=1 case \ZamoGold, $S_{LR}$ is given by \slr\ for all
scattering processes, independent of the charge. The resulting TBA is given by
the diagram
\bigskip
\noindent\centerline{
\rlap{\raise1.05cm\hbox{$L$ \raise.1cm\hbox{\vrule width1cm height
.4pt}$R$}}\rlap{\lower.3cm\hbox{$R$ \raise.1cm\hbox{\vrule width1cm height
.4pt}$L$}}
\hbox{\vrule width.4pt height1cm $\hskip 1.2cm$ \vrule width.4pt height1cm }}
\bigskip
\noindent
In the IR, this TBA system gives $c$=1, as required.  In the UV limit, we have
$c$=3 and log-type corrections given by \cuv\ with $k=-1$.  Note that this
diagram fits in nicely with the $\widehat D_{k+4}$ Dynkin diagrams obtained in
the previous section: it is the $k=-1$ member of this series.  In sect.\ 6 we
will also see how this model can be related to $N$=2 sine-Gordon at a coupling
corresponding to $k=-1$.

This allows an intriguing Landau-Ginzburg interpretation of this model.
Plugging $k=-1$ into \Wis\ gives a superpotential of $W=X$. In this case the
Witten index is zero, so it is possible for supersymmetry to be spontaneously
broken. Such a superpotential is trivial if the kinetic term is $\bar X X$,
but the scattering indicates that this theory is certainly interacting.  This
indicates that our kinetic term is not the simple one, an issue we return to
in sect.\ 7.

\subsec{(0,2) supersymmetry}

Models with two right-moving supersymmetries and no left-moving
supersymmetries are of interest to string phenomenologists.  Motivated by
this, we briefly mention that we can couple our massless $N$=2 right-moving
excitations to $N$=0 left moving excitations to obtain a theory with
non-trivial RG flow into a conformal theory with left movers a $N$=0 minimal
model and right movers a $N$=2 minimal model. This naturally associates a
specific $N$=0 model with each $N$=2 model.

Above, we saw that the $N$=2 $S$-matrices were given by the $N$=0 ones
\refs{\ZamoGold ,\FSZii} with an extra $N$=2 piece tensored to $S_{LL}$ and
$S_{RR}$. The $N$=2 right-moving massless excitations can be coupled to
left-moving $N$=0 massless excitations
merely by tensoring the $N$=2 piece only for $S_{RR}$, i.e.
\eqn\zerotwo{S_{RR}=S_{RR}^{N=0} \otimes S_{u,d}^{N=2}
\quad\qquad S_{LL}=S_{LL}^{N=0}
\quad\qquad S_{LR}=S_{LR}^{N=0},}
where $S^{N=2}_{u,d}$ is that displayed in \Smat.  This yields a TBA system
described by the equations \TBAFis\ and \TBAe\ with the diagram corresponding
to removing the two open nodes on the far left (the $e^{\pm i
\alpha _L}$ nodes) of the $\widehat D_{k+4}$ diagram obtained in sect.\ 2.
(Removing just one of the nodes gives a (1,2) theory which flows into the left
movers of a $N$=1 minimal model combined with the right movers of a $N$=2
model).  The $c$-function $c(\alpha _R=0,M\beta)$ flows from $(2k+3)/(k+3)$ in
the UV to the average of $c_L=1-(6/(k+2)(k+3))$ and $c_R=3k/(k+2)$ in the IR
corresponding to the fact that the IR theory is a $N$=0 minimal model for left
movers combined with a $N$=2 minimal model for right movers.  Note also that
inserting $(-1)^{F_R}$ (by setting $\alpha _R=\pi$ in the integral equations)
removes the $R$ node and its outside $\pm F_R$ nodes from the diagram, leaving
just the left-moving $N$=0 massless scattering theory along the entire flow.

It remains to be seen the extent to which these theories make sense and
if they can be used in the construction of (0,2) string vacua.

\newsec{Adding a background field}

Further information about the physics encoded in our exact $S$-matrices
can be extracted by studying the response of the theories to a constant
external background field.  The two conserved charges $F_L$ and $F_R$
can be coupled to two independent background fields $A_L$ and $A_R$,
modifying the hamiltonian to be $H=H_0+A_LF_L+A_RF_R$.  The exact
$S$-matrices can be used to directly calculate the contribution of these
background fields to the energy density.  Because the background fields
have dimension of mass, their strength controls the position of our
theory on its renormalization group trajectory.  In this section we will
focus on the IR and UV limits, corresponding to small and large
background field strengths, respectively.  More detailed information
about the theories away from their UV and IR fixed points will be
discussed in the next section.

First we consider the flow with $k$=1. The effect of background fields with,
say, $A_L$ and $A_R$ positive is to introduce $d_L$ and $d_R$-type particles
(of charge $-\half$) into the ground state. Left movers fill all levels with
rapidity greater than some value $B_L$, while the right movers fill all with
rapidity less than $-B_R$.  Using the earlier work \refs{\FSZii,\sausage}\ we
can write the answer down: the ``dressed'' particle energies solve
\eqn\fore{\e _L(\t )=\half A_L - {M\over 2}e^{-\t} +
\int_{B_L}^{\infty} d\t ^{\p}\phi_{LL}(\t -\t ^{\p})\e _L(\t ^{\p}) +
\int_{-\infty}^{-B_R} d\t ^{\p}\phi_{LR}(\t -\t ^{\p})\e _R(\t ^{\p}),}
and likewise for $\e_R (\t )$, where $$\phi_{ab}(\tht)\equiv -{i\over 2\pi}
{\del\ln{S_{ab}(\tht)}\over\del\tht}.$$
$\phi_{LL}$ follows from the $dd$ scattering in \Smat, while $\phi_{LR}$
follows from \slr:
$$\phi_{LL}(\t)=
\int_{-\infty}^{\infty} {d\omega\over 2\pi} {\cos \omega\t
\over 4 \cosh ^2{\pi\omega\over 2}}
\qquad\quad\phi_{LR}(\t)={1\over \cosh\t}.$$
The ground-state energy density is then given by
\eqn\gsii{{\cal E}(A_L,A_R)= -\int_{B_L}^{\infty}{d\t \over 4\pi}
\ Me^{-\t}
\ \e_L(\t) -\int _{-\infty}^{B_R}{d\t \over 4\pi} Me^{\t}\ \e_R(\t).}
The $B$ are determined by the boundary condition $\ep_L(B_L)=\ep_R(B_R)=0$.

Finding the equations for arbitrary $k$ requires a little more effort: one
must include zero-mass, zero-charge ``pseudoparticles'' to account for the
fact that we can have different kinds of solitons in the vacuum.
After simplification, the answer is given by \fore--\gsii\ where $\phi_{LL}$
and $\phi_{LR}$ are now
$$\phi_{LL}(\t)=
\int_{-\infty}^{\infty} {d\omega\over 2\pi} \cos \omega\t
\left[{1\over 4\cosh^2{\pi\omega\over 2}}+
{\sinh{(k-1)\pi\omega\over 2}
\over 2 \sinh{k\pi\omega\over 2}\cosh{\pi\omega\over 2}}
\right]$$
$$\phi_{LR}(\t)=
\int_{-\infty}^{\infty} {d\omega\over 2\pi} \cos \omega\t
{\sinh{\pi\omega\over 2}\over 2 \sinh{k\pi\omega\over 2}
\cosh{\pi\omega\over 2}}.$$

These equations can be solved in the $M/A \to 0$ (UV) and $M/A\to\infty$ (IR)
limits.  In the IR limit, the last term in \fore\ is negligible and the
equations for $\ep_L$ and $\ep_R$ decouple.  In the UV limit, one follows the
method of \sausage, yielding
\eqn\lims{\eqalign{{\cal E}_{IR}(A_L,A_R)&
={k\over k+2}{A_L^2+A_R^2\over 4\pi}\cr
{\cal E}_{UV}(A_L,A_R)&=
\left(1-{2k\over(k+2)(k+4)}\right){A_L^2+A_R^2 \over 4\pi}
+{8\over k+4}{A_L A_R\over 4\pi}.\cr}}
The IR limit is in perfect agreement with the general result of \pkmag\ that
the coefficient of $A_L^2+A_R^2$ at a critical point must be $c/12\pi$.  The
UV limit is not that of a $c$=3 CFT, but this is not surprising given the
subtleties of the UV limit mentioned at the end of sect.\ 2. We discuss this
further in sect.\ 6.

\newsec{The UV limit and the $N$=2 sine-Gordon model}

The UV limit of our LG flow theories is related to that of a massive
integrable theory, the $N$=2 super-sine-Gordon model.  The $S$-matrix of $N$=2
sine-Gordon is a tensor product of the $N$=0 sine-Gordon $S$-matrix with the
basic $N$=2 $S$-matrix \Smat\ \ku.  The value of the sG coupling constant which
is related to our $k$-th $N$=2 LG flow theory corresponds to taking the $N$=0
sine-Gordon $S$-matrix at $\beta ^2_{N=0}= 8\pi (k+2)/(k+3)$.  The $N$=2
sine-Gordon TBA system for integer $k\ge 0$ is described by the diagram \pkii
\bigskip
\noindent
\centerline{
\hbox{\rlap{\raise28pt\hbox{$\hskip.25cm e^{i\alpha_F}
\bigcirc\hskip 3.7cm\bigcirc\hskip.1cm e^{i(k+2)\alpha _T}$}}
\rlap{\lower27pt\hbox{$e^{-i\alpha_F}
\bigcirc\hskip 3.7cm\bigcirc\hskip.1cm e^{-i(k+2)\alpha _T}$}}
\rlap{\raise14pt\hbox{$\hskip.9cm\Big\backslash\hskip3.6cm\Big/$
}}
\rlap{\lower14pt\hbox{$\hskip.85cm\Big/\hskip3.55cm\Big\backslash$
}}
$\hskip.8cm${\raise1pt\hbox{$\bigotimes$}}------$\bigcirc$-- --
--
--$\bigcirc$------$\bigcirc$ }}
\bigskip
\noindent
where there are $k+4$ open nodes in all (including the ones at the ends).  As
before $\nu_a=0$ for the open nodes, while for the node $\bigotimes$, it is
$M\cosh\t$.  The phases correspond to turning on $e^{i(\alpha _FF +\alpha _T
T)}$ in the partition function, where $F$ is the conserved fermion number
charge and $T$ is the conserved topological charge counting solitons minus
antisolitons.  Comparing this diagram to the one obtained for our LG flow
theories reveals an obvious similarity: they are both $\widehat D_{k+4}$
Dynkin diagrams.  We also see obvious differences corresponding to the fact
that the LG flow theory has massless excitations and a non-trivial IR fixed
point at $c_{IR}=3k/(k+2)$ whereas the $N$=2 sine-Gordon theory is massive and
must flow to $c_{IR}=0$.  Also, the conserved charges in the LG flow theory
are left and right fermion number whereas the conserved charges of $N$=2
sine-Gordon are the total fermion number $F$ and the topological charge $T$.
We will see that the two theories, while very different in the IR, are deeply
related in the UV.

First we note that both theories have $c$=3 in the UV.  In fact, the
$N$=2 sine-Gordon theory also has precisely the same following term in
\cuv\ (so it too only asymptotically approaches its UV fixed
point).  The fact that this second term agrees for the two theories is a
consequence of the fact that they both have the same diagram.  The
$c$-functions for the two models of course are not the same for all $M\beta$;
numerically, we find that the leading difference between the two fits nicely
to the power $(M\beta )^{4/(k+2)}$.  The similarity between the two theories
in the UV limit remains if we turn on $\alpha _L$=$\alpha _R$=$\alpha$ in the
LG flow theory and $\alpha _F$=$(k+2)\alpha _T$=$\alpha$ in the $N$=2 sG
theory, as can be seen by comparing their diagrams.

We can make this connection even stronger by comparing the two
theories in the presence of a background field.  As discussed in the previous
section we can compute the ground-state energy of the LG theory when the
conserved charges $F_L$ and $F_R$ are coupled to constant background fields
$A_L$ and $A_R$.  Likewise, we can consider the $N$=2 sG theory when the
conserved charges $F$ and $T$ are coupled to background fields $A_F$ and
$A_T$.  We will, in particular, compare the LG flow theory with
$A_L$=$A_R$=$A$ to the $N$=2 sG theory with $A_F$=$(k+2)A_T$=$A$.  Our result
will be the following relation between the two theories.  For the LG theory
the energy density is of the form
\eqn\pows{{\cal E}(A)= b_0 A^2 + \sum_{n=1}^{\infty} (-1)^n b_n
\left({M\over A}\right)^{4n/(k+4)},}
while for the $N$=2 sG theory the energy has exactly the same expansion with
the same $b_n$, only without the $(-1)^n$.  This result indicates that the
models have the same UV fixed point (as $M\to 0$), and that the perturbing
operators are related in a simple manner.  This relation is analogous to
perturbation by $\pm\Phi_{1,3}$ in the $N$=0 minimal models: with one sign the
perturbation flows to the next minimal model and has a massless spectrum,
while the other gives a massive field theory. A power series around the common
UV fixed point has the same behavior, with alternating signs in one case but
not the other.  Another analogous example is the $CP^1$ sigma model where the
theories with $\Theta =\pi$ and $\Theta =0$ are so related \sausage .

We start with the $N$=2 sine-Gordon model.  For our range of the sG coupling
the spectrum consists of four solitons with fermion number $F$ and topological
charge $T$ given by $(F,T)$=$(\pm \half ,\pm 1)$, for the four different sign
choices.  As discussed above, our result requires coupling the background
fields to the two conserved charges as $A_F$=$(k+2)A_T$=$A$.  It so happens
that this ratio of the background fields is special.  For this ratio, with
$A>0$, only the solitons with $(F,T)$=$(-\half, -1)$ appear in the vacuum and
the relevant equations follow immediately from their $S$-matrix element ---
there is no need for pseudoparticles.  This is seen as follows.  Suppose we
had $A_F>0$ and $A_T$=0.  The ground state then fills with the solitons with
$(F,T)$=$(-\half , 1)$ and $(-\half, -1)$.  Now start to increase $A_T$.  At
some special value of $A_T$ it will no longer be energetically favorable to
have the $(-\half , 1)$ particles in the vacuum.  For this ratio of $A_T$ to
$A_F$ there will only be $(-\half, -1)$ states in the vacuum, and from their
single $S$-matrix element we immediately obtain
\eqn\foresg{\e(\t )=\half A_F +A_T  - M\cosh \t  +
\int_{-B}^{B} d\t ^{\p}\Phi(\t -\t ^{\p})\e (\t ^{\p}),}
where
\eqn\ker{\Phi(\t)=
\int_{-\infty}^{\infty} {d\omega\over 2\pi} \cos \omega\t
\left[{1\over 4\cosh^2{\pi\omega\over 2}}+
{\sinh{(k+1)\pi\omega\over 2}
\over 2 \sinh{(k+2)\pi\omega\over 2}\cosh{\pi\omega\over 2}}
\right],}
the first term resulting from the $N$=2 part of the $S$-matrix and the second
from the sine-Gordon part.  The energy density is here
\eqn\gssg{{\cal E}(A)= -{m\over 2\pi}\int_{-B}^{B}d\t\ \cosh\t
\ \e(\t)}
with $B$ determined by the boundary condition $\ep(\pm B)=0$.
In the UV limit, it is straightforward to
show that \sausage
\eqn\gsiv{{\cal E}(A\rightarrow \infty )
= -{q^2A^2\over 2 \pi}{1\over 1-\tilde\Phi(0)} }
where $\tilde\Phi$ is the Fourier transform of $\Phi$. Thus for our case
\eqn\uvmag{{\cal E}(A\rightarrow \infty )= -{4(k+2)\over k+4}
{(\half A_F +A_T)^2\over 2 \pi}}

To prove that this special ratio is $A_F=(k+2)A_T$, we consider two other
choices of the background fields by setting either $A_T$ or $A_F$ to zero.
For $A_F>0$ and $A_T$=0, the vacuum fills with both $(F,T)$= $(-\half,1)$ and
$(-\half, -1)$ particles and so relation \foresg\ with kernel \ker\ doesn't
hold.  Since these particles do not scatter diagonally (their $S$-matrix
involves the sG $S$-matrix), we need pseudoparticles just as in the TBA.
Because the equations are a generalization of \foresg\ and are linear in the
$\e_a (\t)$, we can simplify the expressions and remove the pseudoparticles.
Similar considerations can be applied to the case of $A_T>0$ and $A_F$=0.  For
these two cases we end up with the same equation \foresg , with $A_T$ or $A_F$
appropriately zero, and with the kernel $\Phi$ replaced by, respectively
\eqn\fandt{\eqalign{
\Phi_F(\t)=&
\int_{-\infty}^{\infty} {d\omega\over 2\pi} \cos \omega\t
\left[1-
{\cosh{(k+3)\pi\omega\over 2}
\over 4 \cosh{(k+1)\pi\omega\over 2}\cosh^2{\pi\omega\over 2}}
\right],\cr
\Phi_T(\t)=&
\int_{-\infty}^{\infty} {d\omega\over 2\pi} \cos \omega\t
\left[1-
{\cosh{(k+3)\pi\omega\over 2} \sinh{\pi\omega\over 2}
\over 2 \sinh{(k+2)\pi\omega\over 2}\cosh^2{\pi\omega\over 2}}
\right]\cr}.}
Using $\Phi _F$ and $\Phi _T$ in \gsiv\ we obtain
\eqn\lims{ {\cal E}(A_F\to\infty, A_T=0)= -{A_F^2\over 2\pi}, \quad
 {\cal E}(A_F=0, A_T\to\infty)=-{2(k+2)}{ A_T^2\over 2\pi}.}
These results are in agreement with the general results of \pkmag; the first
gives $c$=3 and the second also shows that the $\beta _{N=2}$ in the
superpotential $W=\cos \half \beta _{N=2}X$ is related to the parameter $k$ in
the $S$-matrix by $\beta _{N=2}^2 =8\pi (k+2)$, a relation which was obtained
in \ku\ by quantum-group symmetry and which can be viewed as a reflection of
$N$=2 nonrenormalization.  The discussion in \pkmag\ also requires that
$${\cal E}(A_F\to \infty, A_T\to \infty)= -{A_F^2\over 2\pi} -
{2(k+2)}{ A_T^2\over 2\pi}$$
because the two currents are independent (this also follows from analyticity
because symmetry rules out a cross term $A_FA_T$).  Comparing this to equation
\uvmag\ obtained above using the kernel \ker , we see that kernel \ker\ is
valid when $A_F=(k+2)A_T$.

We now compare \foresg , with $A_F$=$(k+2)A_T$=$A$ and with
kernel \ker , to the LG flow equations \fore\ with $A_L$=$A_R$=$A$.
First, in the $A\to \infty$ UV limit they both give
$${\cal E}(A\to \infty)=-{k+4\over k+2}{A^2\over 2\pi}.$$
For the $N$=2 sG theory, as discussed above, this result is expected from the
considerations of \pkmag\ for coupling to our peculiar combination of fermion
number and topological charge.  For the LG flow theory, we are coupling to
$F_L+F_R$ which might be interpreted as the total fermion number $F$ (at least
in the IR or on pure left-moving or pure right-moving states).  This result
looks odd in light of the discussion in \pkmag .  Once again, we are seeing
that the UV limit of the LG flow theory is subtle.  Perhaps the analogy with
$N$=2 sG where some topological charge is mixed in with the fermion number
will be useful for better understanding the UV limit of the LG flow theory.
We also note that, if we continue $A$ to imaginary $A=i\alpha /\beta$, we make
contact with the LG theory with $\alpha _L$=$\alpha _R$=$\alpha$ and the $N$=2
sG theory with $\alpha _F$=$(k+2)\alpha _T$=$\alpha$.  These two UV limits
coincide here as in the TBA, but we do not see the peculiar transition
\cuvalpha\ which the TBA Casimir energy exhibits.  This suggests that this
particular aspect of the UV limit is due to the presence of a level crossing
which occurs in the finite-size TBA but not in the infinite-volume
background-field calculation.

The more detailed result \pows\ is obtained by analyzing our two sets of
background field energy equations using a generalized Weiner-Hopf technique
\jnw . Since this has been described in detail for several very similar models
\refs{\hasen,\sausage,\FSZii}, we do not present the full calculation here.
Instead, we will explain how to extract the relevant information from the
kernels. The technique relies on the usual Weiner-Hopf trick of dividing the
Fourier transforms of the kernels into a product of two pieces, the first of
which has no poles or zeroes in the lower half plane and the second none in
the upper half plane.  Writing $1-\tilde\Phi(\omega)=
1/K_+(\omega)K_-(\omega)$, expressions of the form
\eqn\contour{\oint
{f(\omega)\over (\omega -i)^2} g(\omega) e^{2i\omega B} {d\omega\over 2\pi
i};\qquad\qquad g(\omega)\equiv{K_+(\omega)\over K_-(\omega)}}
occur regularly in the analysis; the contour covers the upper half plane. The
function $f(\omega)$ is different depending on where we are in the analysis,
but it is analytic in the upper half plane. For our case of $N$=2 sG the
kernel \ker\ gives
\eqn\kerft{1-\tilde\Phi=
{\sinh{(k+4)\pi\omega\over 2}
\over 4 \sinh{(k+2)\pi\omega\over 2}\cosh^2{\pi\omega\over 2}}.}
Notice that the double pole in $\Phi$ at $\omega =i$ becomes a double zero in
$K_+$ which cancels the explicit double pole appearing in \contour .
Ordinarily, such a pole results in a bulk term proportional to $M^2$; this is
a nice check, because such terms do not appear in supersymmetric theories.
The poles in the contour are the zeros of \kerft , which are at
$\omega=2ni/(k+4)$. Thus \contour\ can be written as a series in
$\exp(-4B/(k+4))$. In particular, the boundary condition results in an
equation
$${M\over A}e^B = const  + \sum_n f_n g_n e^{-4nB/(k+4)}$$
where $f_n$ and $g_n$ are the residues of $f(\omega)/(\omega-i)^2$ and
$g(\omega)$, respectively. The $f_n$ themselves also obey an equation of this
form, so for large $A/M$, we can write $e^B$ and $f_n$ each as a series in
$(A/M)^{-4/(k+4)}$. The energy is also given by a term like \contour, so it
too must be a series in $(A/M)^{-4/(k+4)}$.

For the $A_T$=0 or $A_F$=0 cases, the kernels \fandt\ result in power-series
expansions with different exponents. This indicates that there is not just one
perturbing operator in the model: the different background fields isolate
different operators.

This gives the result \pows\ for $N$=2 sine-Gordon. For the LG flow, we must
first rewrite the equations \fore\ in Weiner-Hopf form.  The general result is
that
$$ \eqalign{qA\to&q (1+{\tilde\phi_{LR}(0)\over 1-\tilde\phi_{LL}(0)})A
\cr
{1\over K_+(\omega)K_-(\omega)}=&1-\tilde\phi_{LL}-{\tilde\phi_{LR}^2\over
1-\tilde\phi_{LL}}\cr g(\omega)=&{K_+(\omega)\over K_-(\omega)}
{\tilde\phi_{LR}\over 1-\tilde\phi_{LL}}.\cr}$$
The $K_+$ and $K_-$ obtained from the LG flow kernels discussed in sect.\ 5
are exactly the same as the ones obtained above for $N$=2 sG. The extra piece
in the above expression for $g(\omega)$ is
$\sinh\pi\omega/\sinh(\pi\omega(k+2)/2)$ here.  It results in no additional
poles in the contour because of the zeros in $g$; its only effect is to change
$g_n$ to $(-1)^n g_n$. Since the $f_n$ above are not changed, we then find
that the series for the LG flow is exactly the same as in $N$=2 sine-Gordon,
except that the signs of every other term are different.  This is what we set
out to prove, and shows that these two models are mysteriously deeply related.

The goldstino $S$-matrix discussed in sect.\ 4 can be related to the
sine-Gordon model at $k=-1$, giving further evidence that the flow in this
case has a superpotential $W=X$. At $k=-1$, the $N$=0 part of the $N$=2 sG
$S$-matrix becomes trivial, leaving the only non-trivial scattering in the
$N$=2 labels. The TBA diagram is given by
\bigskip
\noindent\centerline{
\rlap{\raise1.1cm\hbox{$\bigcirc$\raise.1cm\hbox{\vrule width1cm height
.4pt}$\bigotimes$}}\rlap{\lower.25cm\hbox{$\bigotimes$\raise.1cm\hbox{\vrule
width1cm height .4pt}$\bigcirc$}}
\hbox{$\hskip.1cm$\vrule width.4pt height1cm
$\hskip 1.2cm$ \vrule width.4pt height1cm }}
\bigskip
\noindent
and the log term is that of \cuv\ with $k=-1$, as it is for the Goldstinos.
The $N$=2 sG background-field calculation is covered by the previous
calculation with $k= -1$, while for the Goldstinos it is given by
\fore--\gsii\ with
$$\phi_{LL}(\t)= 0
\qquad\qquad\phi_{LR}={1\over \cosh\t}.$$
It is simple to verify that the two expansions are of the form \pows\ with an
extra ``bulk'' term for the Goldstinos because the pole at $\omega=i$ is not
cancelled; this is allowed because the supersymmetry is spontaneously broken.

We note that the $k$=0 case of the LG flow (corresponding to $W=X^2$) is
massive with a trivial fixed point. It seems to be not just related but
actually identical to the $N$=2 sine-Gordon theory at $k$=0.  We don't know
why this is so, but this may be helpful in understanding the relation between
the two theories for all $k$.

\newsec{Questions and conclusions}

We have seen that the quantitative results from our ``LG flow''
scattering theory match a variety of quantitative and qualitative
expectations for the LG theory
\LGaction\ with the superpotential \Wis .  One might hope to be able to do
better by connecting the results to some more detailed aspects of the LG
theory.  For example, in the UV limit we might be able to make contact with a
perturbative analysis of the action \LGaction\ with some particular kinetic
term $K$ and with the superpotential \Wis\ small.  As discussed in the
introduction, this is hard; there are difficultities in regulating and
analyzing the theory with the superpotential \Wis.  The basic problem is that
the bosonic component of $X$ appears directly in the action and, with the
standard standard $K=X\bar X$ kinetic term, the boson fluctuates too much.  We
can handle derivatives and exponentials of the boson, but the logarithms in
the boson-boson correlation functions make it difficult to work with the boson
itself.

We saw that the Goldstino case $k=-1$ is described by the superpotential $X$.
If the kinetic term were $\bar XX$, even if only in the UV, this model would
be trivial all along the flow. Since our $S$-matrix is certainly that of an
interacting theory, we are motivated to try a different kinetic term. We can
hope that this will also make the boson better behaved.  We consider taking
the boson to live on some sigma model with metric $G_{x\bar x}=\partial
_x\partial _{\bar x}K$.  In order to have this theory give the right elliptic
genus the sigma model should be topologically equivalent to the plane.  A
choice of metric which looks promising in the UV is the cigar
\eqn\metric{G_{x\bar x}=(al^{b\over 4\pi}+bx\bar x)^{-1}}
where $a$ and $b$ are constants and $l$ is the RG length scale, say
$l=M\beta$, so $l\to 0$ in the UV.  The kinetic term corresponding to this
metric is a dilogarithm function. Unlike the black hole of \withole, we do not
have a dilaton so our metric has the above nontrivial RG flow; it is a
solution of the flow equation discussed in \sausage\ when the superpotential
is turned off.  Treating the superpotential as a small perturbation (a
``tachyon condensate''), there will be order $g^4$ corrections to the RG flow
of this metric. The superpotential \Wis\ scales to zero in the UV limit
provided $b(k+2)^2<2$.

It remains to be seen if one can develop a sensible perturbation theory using
the cigar metric with the superpotential \Wis . One way of finding the
appropriate kinetic terms for these theories may be to study classical
integrable equations, along the lines of \evanshollo.  We simply note that it
appears possible for this theory to reproduce one aspect of our calculation,
the asymptotic $1/\log ^2M\beta$ behavior \cuv.  As in \rAlZiv\ these terms
come from the fields which are (nearly) constant in the $\beta$ cycle of the
torus.  For $\alpha _L$ and $\alpha _R$ zero, the fermions are antiperiodic in
this cycle so we can neglect them.  The contribution of the boson constant
modes to the ground state energy reduces to a quantum mechanics problem in the
constant mode coordinates $x_0$ and $\bar x_0$.  It seems plausible that this
gives the behavior \cuv, since in the $N$=0 sausage sigma model the identical
terms arise \sausage.  The metric \metric\ is also given qualitative support
by some recent results \VV , where similar metrics are considered with the
superpotential \Wis\ in a Landau-Ginzburg description of some
black-hole-motivated sigma models, along the lines of \MukVaf. This opens up
the intriguing possibility that our exact LG flow scattering theories are
related to the world-sheet physics of 2D black holes.

Understanding the ultraviolet limit of these theories better will almost
certainly shed more light on the situation. Indeed, we have a number of
unanswered questions here:
Can we see the connection to $N$=2 sine-Gordon theory
directly from the Lagrangians? Why in this correspondence is
topological charge mixed in with fermion number?
Why is the $k=0$ model identical in both these cases?

Our exact massless scattering theories provide quantitative information which,
as we have seen, agree with results and expectations for the $N$=2 LG flows.
On the other hand, the program of describing RG flows to nontrivial IR CFTs in
terms of integrable scattering theories of massless excitations is very new
and not yet completely understood.  For example, it is not known in detail how
in the IR the decoupled left and right massless scattering theories are
``equivalent'' to the usual descriptions of the IR CFTs.  For example, the
left and right scattering theories in the IR limit are totally decoupled, but
we know that the left and right CFTs are not.  While we know much about these
massless excitations (e.g.\ their exact $S$-matrices and the free energy), one
wonders what these excitations ``really are''. Are they real particles, or are
they just a way of encoding exact information like the free energy?  We
hope that these $N$=2 examples will shed some light on these general issues.
In particular, perhaps the $N$=2 superconformal algebra obtained in \witlg\
acting on purely left-moving or right-moving states will allow for a better
understanding as to what these massless excitations really are.

\centerline{\bf Acknowledgements}

We would like to thank V. Fateev, C. Vafa and A. Zamolodchikov for very
helpful conversations.  P.F.\ was supported by DOE grant DEAC02-89ER-40509,
K.I. by DE-FG05-90ER40559.

\listrefs

\end